\documentclass[aps,prfluids,10pt,a4paper,twocolumn,superscriptaddress,balancelastpage,showpacs,reprint,notitlepage]{revtex4-2}

\usepackage[english]{babel}
\usepackage[T1]{fontenc}
\usepackage{endnotes}
\usepackage{amssymb,amsmath,amsfonts}
\usepackage{textcomp}
\usepackage{graphics,color}
\usepackage[colorlinks,citecolor=blue,urlcolor=blue,bookmarks=false,hypertexnames=true]{hyperref} 

\usepackage{graphicx}

\usepackage{float}
\usepackage[export]{adjustbox}
\usepackage[abs]{overpic}
\usepackage{xcolor,varwidth}

\usepackage{layouts}

\usepackage{DejaVuSans}

\begin{document}

\title{Anisotropic diffusion of ellipsoidal tracers in microswimmer suspensions}

\author{Henrik Nordanger}
\affiliation{Division of Physical Chemistry, Lund University, 221 00 Lund, Sweden}

\author{Alexander Morozov}
\affiliation{SUPA, School of Physics and Astronomy, The University of Edinburgh, James Clerk Maxwell Building, Peter Guthrie Tait Road, Edinburgh, EH9 3FD, United Kingdom}

\author{Joakim Stenhammar}
\email{joakim.stenhammar@fkem1.lu.se}
\affiliation{Division of Physical Chemistry, Lund University, 221 00 Lund, Sweden}



\date{\today}

\begin{abstract}
\noindent Tracer particles immersed in suspensions of biological microswimmers such as \emph{E. coli} or \emph{C. reinhardtii} display phenomena unseen in conventional equilibrium systems, including strongly enhanced diffusivity relative to the Brownian value and non-Gaussian displacement statistics. In dilute, 3-dimensional suspensions, these phenomena have typically been explained by the hydrodynamic advection of point tracers by isolated microswimmers, while, at higher concentrations, correlations between pusher microswimmers such as \emph{E. coli} can increase the effective diffusivity even further. Anisotropic tracers in active suspensions can be expected to exhibit even more complex behaviour than spherical ones, due to the presence of a nontrivial translation-rotation coupling. Using large-scale lattice Boltzmann simulations of model microswimmers described by extended force dipoles, we study the motion of ellipsoidal point tracers immersed in 3-dimensional microswimmer suspensions. We find that the rotational diffusivity of tracers is much less affected by swimmer-swimmer correlations than the translational diffusivity. We furthermore study the anisotropic translational diffusion in the particle frame and find that, in pusher suspensions, the diffusivity along the ellipsoid major axis is higher than in the direction perpendicular to it, albeit with a smaller ratio than for Brownian diffusion. Thus, we find that far field hydrodynamics cannot account for the anomalous coupling between translation and rotation observed in experiments, as was recently proposed. Finally, we study the probability distributions (PDFs) of translational and rotational displacements. In accordance with experimental observations, for short observation times we observe strongly non-Gaussian PDFs that collapse when rescaled with their variance, which we attribute to the ballistic nature of tracer motion at short times. 
\end{abstract}
\maketitle

\section{Introduction}

Active transport of particles is of importance in many biological contexts, such as intracellular transport~\citep{Koslover:PhysBiol:2020}, absorption of nutrients in intestines~\citep{Janssen:FoodFunct:2015} and by microorganisms~\citep{Koehl:PRL:2013}, and possibly mass transport in oceans~\citep{Katija:ExpBiol:2012}. It also has a more fundamental relevance, as tracers can be used to probe the nature of non-equilibrium fluctuations in active systems, in particular in relation to (equilibrium) Brownian motion~\citep{Volpe:PRE:2016}. Arguably, the simplest model system for studying tracer diffusion in such ``active baths'' is a spherical colloidal particle immersed in a dilute suspension of biological microswimmers, such as \emph{E. coli}~\citep{Wu_Libchaber:PRL:2000,Drescher1,Jepson1,Kim1,Koumakis1,Mino1,Mino2,Patteson1,Peng1,Semeraro1} or \emph{C. reinhardtii}~\citep{Leptos1,Ortlieb1,Yang1,Eremin:2021}. A generic feature of tracer motion in these systems is that the displacement is characterized by ballistic motion at short times (approximately $0.02 - 2$ seconds~\citep{Patteson1}) and diffusive motion at long times, although with a significantly increased translational diffusion coefficient $D_T$ compared to the corresponding Brownian one. Furthermore, in the concentration regime where swimmer-swimmer correlations are sufficiently weak, $D_T$ displays a linear scaling with swimmer density~\citep{Mino1,Jepson1,Leptos1}. For dilute, 3-dimensional microswimmer suspensions, this enhanced tracer diffusion has been rationalised from the superposition of independent hydrodynamic swimmer-tracer scattering events due to the swimmers’ long-ranged dipolar flow fields~\citep{Morozov1,Delmotte1,Dunkel2,Jepson1,Lin1,Thiffeault2,Yeomans:JFM:2013}, while, at much higher densities, for finite-size particles, or in confined geometries, direct swimmer-tracer collisions~\citep{Lagarde1,Brady:PRE:2017,Xu:ChinJCP:2021} and near-field hydrodynamic effects~\citep{Dyer:PoF:2021,Patteson1} also become significant. Furthermore, for the case where microswimmers are significantly larger than the tracer particles, so-called tracer entrainment can occur, leading to isolated, very large displacements of individual tracers and a qualitative change in the displacement statistics~\citep{Jeanneret1,Yeomans:PRFluids:2017,Yeomans:JFM:2013}. Another well-studied non-equilibrium feature of tracer dynamics in dilute microswimmer suspensions is the probability distribution function (PDF) of tracer displacements within a fixed time window. For Brownian diffusion, these PDFs are Gaussian, while in biological microswimmer suspensions they become strongly non-Gaussian at low swimmer densities and short observation times~\citep{Leptos1,Ortlieb1,Kurtuldu:PNAS:2011}. This effect is explained by the fact that tracer displacements result from just a small number of swimmer-tracer scattering events, so that the central limit theorem does not apply~\citep{Kanazawa1,Zaid1,Thiffeault1}, and has also been reproduced in computational models of microswimmers~\citep{Thiffeault1,Lin1,Delmotte1}. 

\begin{figure}[ht]
\centering

\includegraphics[width=75mm, clip, trim=0cm 0cm 0cm 0cm]{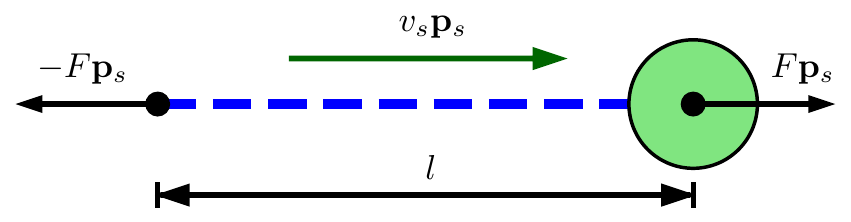}
\caption{Schematic illustration of a pusher-type model swimmer and its parameters.}
\label{fig:model}
\end{figure}

Even though the enhanced tracer diffusion at sufficiently low microswimmer densities can be approximated as arising from independent swimmer-tracer scattering events, swimmer-swimmer correlations will strictly affect the enhanced diffusion for any finite microswimmer concentration~\citep{Stenhammar1,Morozov:PRX:2020}. When such correlations are taken into account, the pusher-puller symmetry and the linear scaling of $D_T$ with swimmer density are broken: In puller suspensions, such as \emph{C. reinhardtii}, the scaling progresses slower than linearly, while in pusher suspensions, such as \emph{E. coli}, the increase becomes superlinear~\citep{Stenhammar1,Qian1,Morozov:PRX:2020,Krishnamurthy1}. At a well-defined critical density of pushers, $D_T$ diverges, corresponding to the onset of active turbulence, a phenomenon in which the system exhibits large-scale vortices and jets, with fluid velocities surpassing the swimming speed of an individual swimmer. ~\citep{Stenhammar1,Bardfalvy:SoftMatter:2019,Koch:AnnuRev:2011,Saintillan1}.

Going beyond the case of spherical tracers, a few recent experimental studies have focused on the dynamics of \emph{anisotropic} passive particles in the form of ellipsoids~\citep{Peng1,Yang1} or dumbbells~\citep{Eremin:2021} immersed in active suspensions consisting of swimming bacteria or algae confined to liquid films. In addition to the enhanced overall diffusion seen for spherical tracers, these studies found significantly increased rotational diffusion coefficients $D_R$. More surprisingly, they found that ellipsoidal tracers display a qualitatively anomalous anisotropic diffusion compared to the Brownian case, in that the ratio $D_{\parallel}/D_{\perp}$ of diffusion coefficients parallel and perpendicular to the particle major axis is below unity in the case of pusher-type swimmers (\emph{E. coli}) at high densities~\citep{Peng1}. This contrasts with the Brownian diffusion of ellipsoids in bulk suspensions, where $D_{\parallel}/D_{\perp} \rightarrow 2$ for large aspect ratios due to their anisotropic friction~\citep{Dhont:Dynamics,Yodh:Science:2006,Han:JCP:2010}. For puller-type swimmers (\emph{C. reinhardtii}), the same authors instead found that the corresponding ratio remains above unity~\citep{Yang1}. These results were partially rationalised as due to the different symmetries of pusher and puller flow fields in quasi-2D geometries~\citep{Peng1,Yang1}, but their precise origin is still unclear. 

To build a better theoretical understanding of anisotropic tracer dynamics in active suspensions, we will here consider a simple computational model of ellipsoidal tracers immersed in 3-dimensional microswimmer suspensions. In terms of their translational dynamics, tracers are described as point particles, while the orientational tracer dynamics are governed by Jeffery’s equation for extended ellipsoids. The swimmers are modelled as extended force dipoles, which affect the tracers only via their flow fields, which we implement using a lattice Boltzmann framework. The simplicity of the model allows for large-scale particle resolved simulations with $N>10^5$ swimmers and $N = 10^5$ tracers, while accurately capturing the far-field contribution to the tracer dynamics. 

Our main finding is that translational diffusion in the particle frame is anisotropic only for the case of pusher suspensions, and furthermore only at densities around and above the transition to active turbulence. In contrast to experimental results for ellipsoidal tracers in \emph{E. coli} suspensions~\citep{Peng1}, we find that the ratio $D_{\parallel}/D_{\perp}$ is \emph{above} unity, indicating that far-field hydrodynamics cannot account for the anomalous coupling between translation and rotation seen in experiments. 

\begin{figure}[ht]
\centering

\includegraphics[width=75mm, clip, trim=0.0cm 0.0cm 0cm 0cm 0cm]{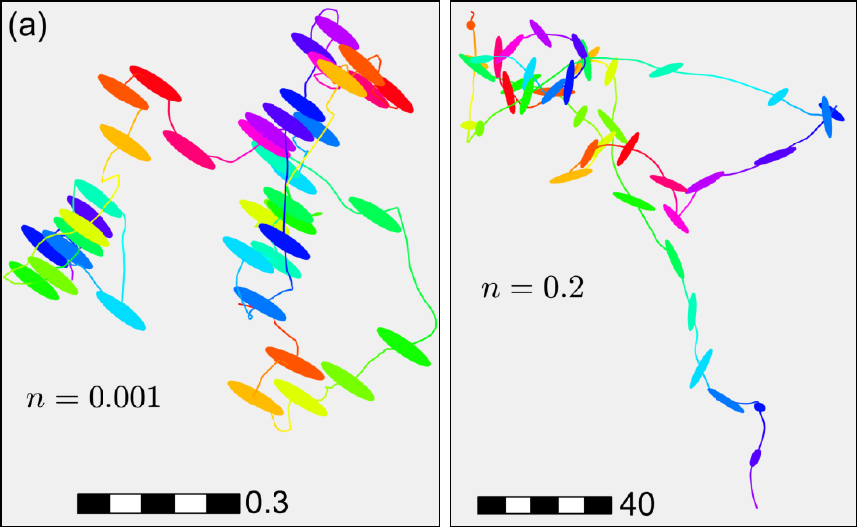}

\includegraphics[width=75mm, clip, trim=0.1cm 0.0cm -0.05cm 0cm]{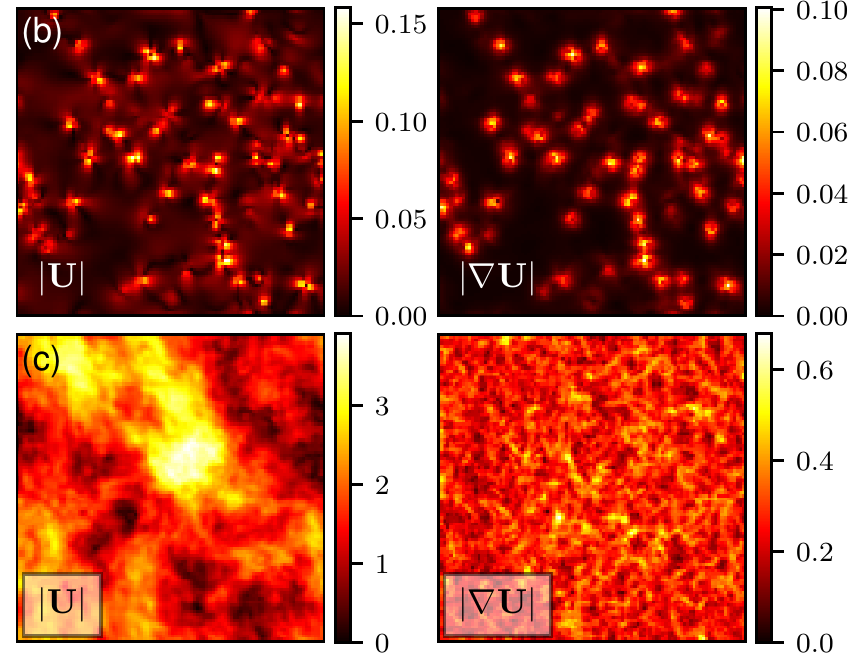}

\caption{(a) Example tracer trajectories in the dilute regime ($n=0.001$, left) and in the turbulent regime ($n=0.2$, right). The length of the scale bars are given in units of the swimmer length $l$. The length of both trajectories is $\Delta t=5\times10^2$, with the changing particle colour denoting different times. 
(bc) Magnitude of the local flow field $\mathbf{U}$ (left) and its gradient $\nabla \mathbf{U}$ (right) in the dilute (b) and turbulent (c) regimes. Both quantities are shown in a slice of the 3-dimensional box. 
} 
\label{fig:overview}
\end{figure}

\section{Model and Method}\label{section:method}

We consider a three-dimensional suspension containing $N_s$ swimmers and $N_t = 10^5$ tracer particles, moving in a box with periodic boundaries. Swimmers are represented by extended force dipoles, where the body and flagella exert two equal and oppositely directed forces $\pm F\mathbf{p}_s$ separated by a distance $l$ on the fluid, where the unit vector $\mathbf{p}_s$ represents the orientation of the swimmer (see Fig.~\ref{fig:model}). Each swimmer is characterized by its dipole strength $\kappa=\pm Fl/\mu$, where $\mu$ is the dynamic viscosity of the fluid, and the sign of $\kappa$ distinguishes pushers from pullers, represented by $\kappa>0$ and $\kappa<0$, respectively.

The time evolution of each swimmer's position $\mathbf{r}_s$ and orientation $\mathbf{p}_s$ is governed by the following equations of motion (EOMs)~\citep{Stenhammar1,Bardfalvy:SoftMatter:2019}:
\begin{align}
\label{eq:rdot} \dot{\mathbf{r}}_s &= v_s\mathbf{p}_s + \mathbf{U}(\mathbf{r}_s) \\
\label{eq:pdot} \dot{\mathbf{p}}_s &=     (\mathbb{I}-\mathbf{p}_s\mathbf{p}_s)\cdot \displaystyle{\frac{\mathbf{U}(\mathbf{r}_s)-\mathbf{U}(\mathbf{r}_s-\mathbf{p}_sl)}{l}},
\end{align}

where $\mathbf{U}(\mathbf{r}_s)$ is the velocity of the fluid at the position of the swimmer's body, $v_s$ is the swimming speed of an individual swimmer, and $\mathbb{I}$ is the unit tensor. As discussed in~\citep{Stenhammar1}, the orientational EOM~\eqref{eq:pdot} describes the dynamics of a flow-aligning swimmer composed of a spherical body of radius $a$ connected by a thin rod of length $l$ to the flagellum, in the limit $a \ll l$ (see Fig.~\ref{fig:model}). In addition to the reorientation caused by hydrodynamic interactions, swimmers undergo run-and-tumble motion, implemented by imposing Poisson distributed random reorientations with an average frequency $\lambda$. In the absence of other microswimmers, these dynamics yield a persistent random walk with the persistence length $v_s/\lambda$. Since the effective diffusion of tracer particles in microswimmer suspensions is independent of $\lambda$ for persistence lengths $v_s / \lambda \gg l$~\citep{Lin1,Pushkin1}, we for simplicity choose to keep $\lambda$ fixed at a value close to that of wild-type \emph{E. coli}, corresponding to $v_s / \lambda l = 5$. 

\begin{figure}[ht!] 
\begin{center}
\includegraphics[width=75mm, clip, trim=4cm 3.14cm 2.2cm 1.64cm]{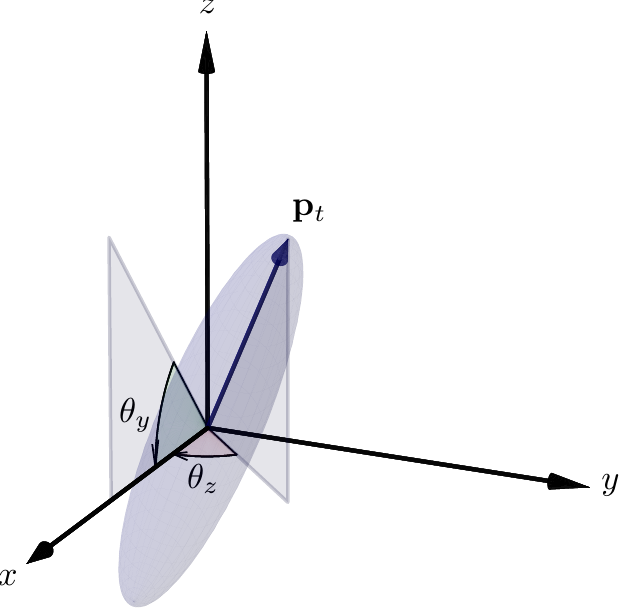}
\caption{Illustration of the rotation scheme used for evaluating particle displacements in the particle coordinate frame.} 
\label{fig:angles}
\end{center}
\end{figure}

The EOMs for the positions $\mathbf{r}_t$ and orientations $\mathbf{p}_t$ of the ellipsoidal point tracers are given by
\begin{align}
    \label{eq:rdot_tr} &\dot{\mathbf{r}}_t=\mathbf{U}(\mathbf{r}_t) \\
    \label{eq:pdot_tr}&\dot{\mathbf{p}}_t=(\mathbb{I}-\mathbf{p}_t\mathbf{p}_t)\cdot
(\beta\mathbf{E}+\mathbf{W})\cdot\mathbf{p}_t,
\end{align}
where Eq.~\eqref{eq:pdot_tr} is Jeffery's equation, describing the reorientation of an ellipsoidal particle in a shear flow. Here, $\beta=(q^2-1)/(q^2+1)$ with $q$ denoting the ellipsoid aspect ratio, and
$\mathbf{E}=(\nabla\mathbf{U}+\nabla\mathbf{U}^\dagger)/2$ and $\mathbf{W}=(\nabla\mathbf{U}-\nabla\mathbf{U}^\dagger)/2$ are, respectively, the rate-of-strain and vorticity tensors evaluated at $\mathbf{r}_t$. For simplicity we set $\beta = 1$, corresponding to infinite aspect ratio, although we also checked that our results only depend weakly on the value of $\beta$ for aspect ratios $q \geq 2$. In this limit, Eq.~\eqref{eq:pdot_tr} simplifies to 
\begin{equation}\label{eq:pdot_tr_infAspect} \dot{\mathbf{p}}_t=(\mathbb{I}-\mathbf{p}_t\mathbf{p}_t)\cdot (\nabla\mathbf{U})\cdot\mathbf{p}_t,
\end{equation}
where the gradient $(\nabla\mathbf{U})\cdot\mathbf{p}_t$ along the tracer orientation was numerically evaluated as the first-order central difference evaluated at the tracer position, similar to the swimmer EOM~\eqref{eq:pdot}. Notably, the translational EOM~\eqref{eq:rdot_tr} is identical to that of a point tracer, so that the anisotropic behavior is fully encoded in the orientational EOM~\eqref{eq:pdot_tr_infAspect}. This is in contrast to Brownian motion of elongated particles, which is intrinsically anisotropic due to the different friction coefficients along the minor and major axes. As Brownian motion is not explicitly included in our simulations, all observed anisotropic behaviour is thus solely due to this intricate translation-rotation coupling.

\begin{figure*}[ht!] 
\centering
\includegraphics[width=160mm]{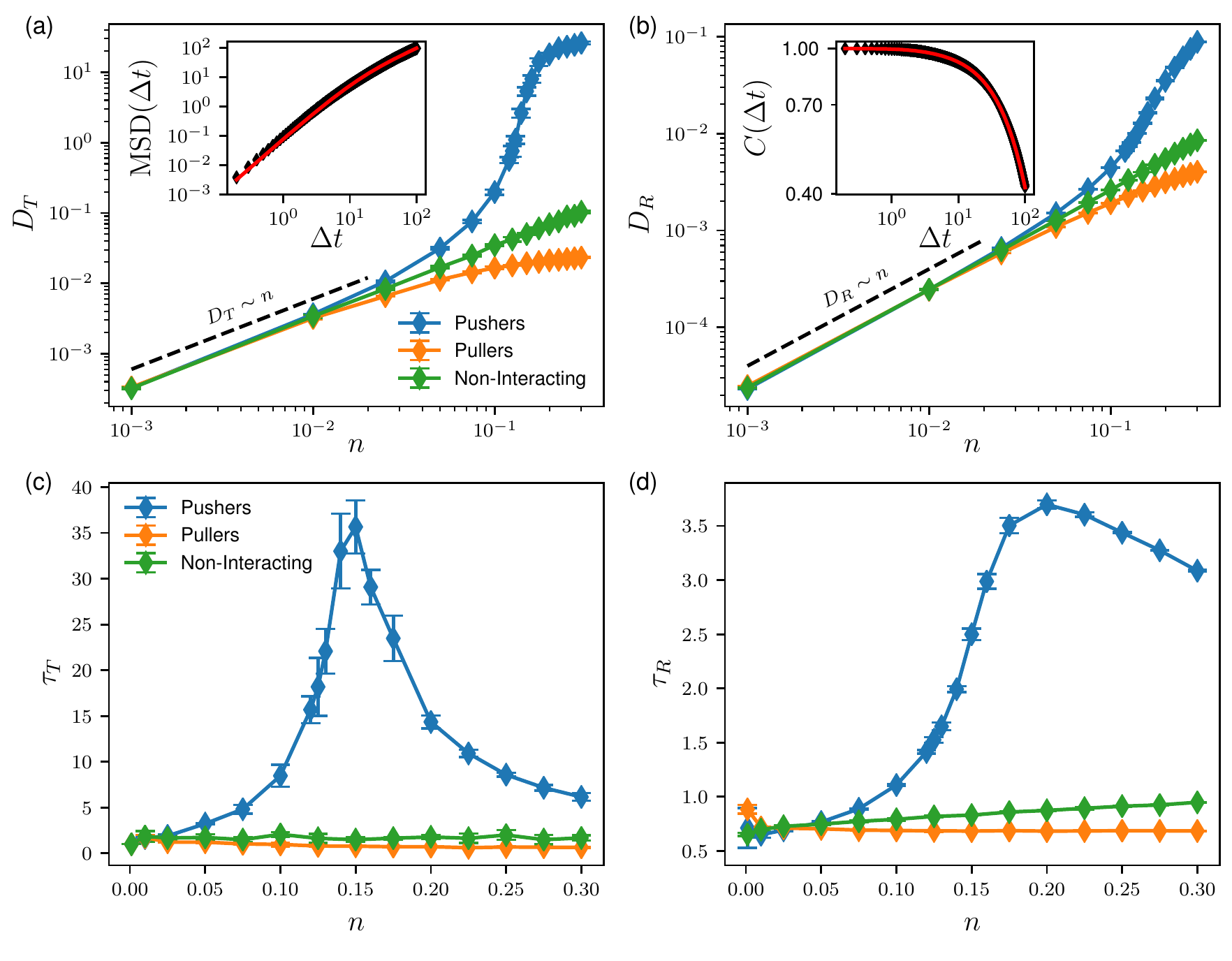}

\caption{\textbf{Translational and rotational diffusion of anisotropic tracers.} Panels (a) and (b) show respectively the  translational ($D_T$) and rotational ($D_R$) diffusion coefficients as functions of the swimmer density $n$, measured in the external coordinate frame. The inset in (a) shows an example of the translational MSD in a pusher suspension at $n=0.1$, with an analytical fit using Eq.~\eqref{eq:MSD_T}. The inset in (b) shows $C(\Delta t)$ in the same suspension, with an analytical fit using Eq.~\eqref{eq:autocorr}. Panels (c) and (d) show the respective correlation times for translational ($\tau_T$) and rotational ($\tau_R$) motion as functions of $n$. Error bars denote estimated standard deviations from at least four separate runs with different initial conditions. 
}
\label{fig:D_Tau}
\end{figure*}

The translational and rotational displacements of tracers were tracked both in the lab frame and by considering a separate, co-moving frame of reference for each individual particle. Our approach is equivalent to that of Ref.~\citep{Yodh:Science:2006}, but extended from two to three dimensions. We start from the tracer position and orientation in the lab frame at time step $t_n$, denoted by $\mathbf{r}_t(t_n)$ and $\mathbf{p}_t(t_n)$, respectively. During a time interval $\delta t=t_n-t_{n-1}$, the particle undergoes a translation $\delta\mathbf{r}_{t,n}=\mathbf{r}_t(t_n)-\mathbf{r}_t(t_{n-1})$, 
which is transformed into its body frame counterpart $\delta\mathbf{\Tilde{r}}_{t,n}$ 
by applying two subsequent rotations (Fig.~\ref{fig:angles}):
\begin{equation}
\label{eq:transformation} \delta\mathbf{\Tilde{r}}_{t,n}=\mathbb{R}_y\cdot \mathbb{R}_z\cdot\delta\mathbf{r}_{t,n},
\end{equation}
with 
\begin{align}
\label{eq:Rotmatrices}&\mathbb{R}_y=\begin{pmatrix}
  \cos{\theta_{y,n}} & 0 & \sin{\theta_{y,n}}\\ 
  0 & 1 & 0\\
  -\sin{\theta_{y,n}} & 0 & \cos{\theta_{y,n}}
\end{pmatrix},\\
 &\mathbb{R}_z=\begin{pmatrix}
  \cos{\theta_{z,n}} & \sin{\theta_{z,n}} & 0\\ 
  -\sin{\theta_{z,n}} & \cos{\theta_{z,n}} & 0\\
  0 & 0 & 1
\end{pmatrix}.
\end{align}
Here, $\theta_{i,n}=[\theta_i(t_{n-1})+\theta_i(t_n)]/2$, $\theta_y$ is the angle between $\mathbf{p}_t$ and the lab $xy$ plane and $\theta_z$ is the azimuthal angle between the $x$-axis and a projection of $\mathbf{p}_t$ onto the $xy$ plane (see Fig.~\ref{fig:angles}). The total body frame displacement of the tracer during a macroscopic time interval $t_n$ is obtained by summing over all displacements $\delta\mathbf{\Tilde{r}}_{t,n}$: 
\begin{equation}
\label{eq:displacement} \mathbf{\Tilde{r}}_t(t_n)=\sum^n_{k=1}\delta \mathbf{\Tilde{r}}_{t,k}.
\end{equation}
Using Eq.~\eqref{eq:displacement}, we consider body frame displacements for trajectories of duration $\Delta t$ via $\Delta \mathbf{\Tilde{r}}_t(\Delta t)=\mathbf{\Tilde{r}}_t(t_0+\Delta t)-\mathbf{\Tilde{r}}_t(t_0)$. The first component of this quantity represents the displacement along the tracer major axis, with the other two components representing the displacement in the plane perpendicular to this axis.

For the evaluation of translational diffusion coefficients, mean-square displacements (MSDs) in the lab frame and in the particle frame were obtained by averaging over all particle trajectories with a given duration $\Delta t$. Due to the persistent character of the fluid flows, the MSDs can be fitted to a persistent random walk with a ballistic regime at short times and a diffusive regime beyond a crossover time $\tau_T$, resulting in~\citep{Maggi:PRL:2014}
\begin{equation}
\mathrm{MSD}(\Delta t) = 6D_T\left[\Delta t-\tau_T(1-\exp{(-\Delta t/\tau_T})\right],
\label{eq:MSD_T}
\end{equation}
where the MSD was obtained by averaging over all tracer particles and all time origins, after removing initial transients due to relaxation of the system to steady state. The rotational diffusion coefficient $D_R$ was calculated by fitting the orientational autocorrelation function $C(\Delta t) \equiv \langle\mathbf{p}(t_0)\cdot\mathbf{p}(t_0+\Delta t)\rangle$ to the approximate expression suggested by ~\citet{Pumir:JStat:2011}, which interpolates between the correct ballistic ($C = \exp\left[-(D_R / \tau_R) \Delta t^2 \right]$) and diffusive ($C = \exp\left[-2D_R \Delta t \right]$) behaviours:
\begin{equation}
C(\Delta t) =\exp\left[-\frac{2D_R\Delta t^2}{\sqrt{4\tau_R^2+\Delta t^2}}\right].
\label{eq:autocorr}
\end{equation}

The numerical evaluation of fluid flows were implemented through the D3Q15 BGK lattice Boltzmann (LB) method as developed by Nash \emph{et al.}~\citep{Nash1,Nash2}, and further discussed in~\citep{Bardfalvy:SoftMatter:2019,Stenhammar1}. We employed a cubic box of size $(100^3)$ lattice points, which is large enough to minimise finite-size effects even in the active turbulence regime~\citep{Bardfalvy:SoftMatter:2019}, and a simulation length of up to $5 \times 10^5$ time steps. 
In terms of LB units, defined by the lattice spacing $\Delta L$ and time step $\Delta t$,  the swimmer parameters were set to $v_s = 10^{-3}$, $F = 1.57\cdot 10^{-3}$, $l = 1$, $\lambda = 2 \cdot 10^{-4}$, and $\mu = 1/6$, where the latter value corresponds to the fluid relaxing to local equilibrium on each timestep. The resulting (non-dimensionalised) value of $\kappa$ corresponds closely to that measured for \emph{E. coli}~\citep{Drescher1,Bardfalvy:SoftMatter:2019}. In the following, rather than using LB units, we will present all results in terms of the swimmer length $l$ and the swimming time scale $l / v_s$. These can in turn be related to physical units by rescaling with the corresponding dipole lengths and swimming speeds of the experimental system in question: for \emph{E. coli}, $l \approx$ 2 $\mu$m, $v_s \approx$ 20 $\mu$m/s, and $F \approx 0.4$~pN~\citep{Drescher1}.

\section{Results and Discussion}

\begin{figure}[h!] 
\begin{center}
\includegraphics[width=80mm]{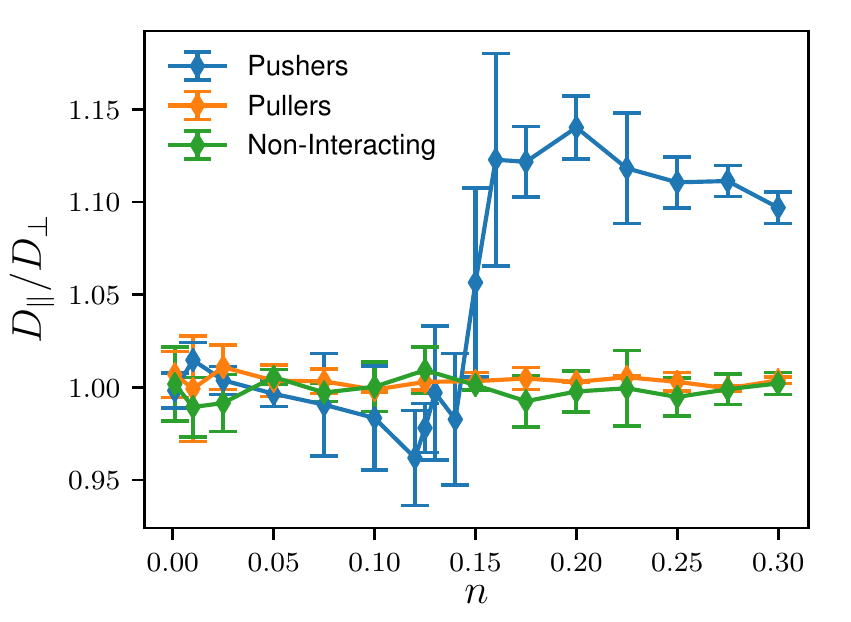}
\caption{\textbf{Anisotropic diffusion in the particle coordinate frame.} Symbols show the ratio $D_{\parallel} / D_{\perp}$ between translational diffusion coefficients along the tracer major and minor axes. Error bars denote estimated standard deviations from at least four separate runs with different initial conditions.} 
\label{fig:anisotropy}
\end{center}
\end{figure}

The model introduced in the previous Section has been extensively used to study the onset of collective motion in dilute suspensions of pusher-like microswimmers~\citep{Stenhammar1,Bardfalvy:SoftMatter:2019,Bardfalvy:PRL:2020}, and its phenomenology is well-understood. Below the threshold value $n_c$ of the number density $n = N/V$ of microswimmers, the suspension remains homogeneous and isotropic while exhibiting significant correlations between microswimmers. As the number density approaches its critical value, the interparticle correlations diverge, culminating in a collective state. The associated velocity fields are correlated across the whole domain and comprise large-scale jets and vortices~\citep{Hohenegger1,Saintillan1,Subramanian1,Bardfalvy:SoftMatter:2019}. In puller-like suspensions, on the other hand, long-range hydrodynamic interactions suppress collective motion and yield suspensions less correlated than their non-interacting counterparts.

In Fig.~\ref{fig:overview}a, we show representative trajectories of a single ellipsoidal tracer immersed in a suspension of pusher microswimmers below  ($n = 0.01$) and above ($n=0.2$) the onset of collective motion (see also movies in~\citep{SM}). For both densities, the tracer exhibits what looks like a ``diffuse-and-tumble'' motion: its center of mass and its orientation perform random walks punctuated by rare significant re-orientation events. The origin of these dynamics lies in the decay law of the velocity fields generated by individual microswimmers. The tracer advection due to a superposition of dipolar fields ($\sim r^{-2}$) generated by the whole suspension is a truly long-range effect in 3D. The tracer rotation, on the other hand, is only marginally long-range since the superposition of the associated velocity gradients ($\sim r^{-3}$) diverges as a logarithm of the system size. 

To further visualise this difference, in Fig.~\ref{fig:overview}b, we plot $\mathbf{U}$ and its gradient in the dilute regime, $n=0.001$. The fluid velocity $\mathbf{U}$ exhibits patches of significant magnitude, though much smaller than $v_s$, embedded in a background of small but finite amplitude. The velocity gradient, on the other hand, has significant values only in the direct vicinity of individual microswimmers. A tracer sampling such fields, performs, effectively, a translational random walk with a constant orientation until a close encounter with a microswimmer (rare in dilute suspensions) changes its orientation significantly. Such sudden reorientations due to close encounters with individual microswimmers can be seen as rotational analogues of entrainment events disscussed above~\citep{Jeanneret1,Pushkin1}.

Above the onset of collective motion ($n=0.2$, Fig.~\ref{fig:overview}c) the fluid velocity is significantly larger than $v_s$ and is correlated across the whole domain. The resulting translational motion is again a random walk with persistence length set by the magnitude of the fluid velocity and the correlation time of the fluid flow, which is signficantly longer than the tumble time $\lambda^{-1}$ of individual microswimmers~\citep{Bardfalvy:SoftMatter:2019,Morozov:PRX:2020}, yielding a significantly enhanced translational diffusivity $D_T$. On the other hand, the velocity gradient is still correlated over short distances only and its maximum values are only somewhat bigger than their dilute counterparts in Fig.~\ref{fig:overview}b. This again implies rare reorientation events interspaced with periods of weak rotational random walk.

To further quantify these observations, in Fig.~\ref{fig:D_Tau} we show the tracer diffusion coefficients $D_T$ and $D_R$ and corresponding crossover times $\tau_T$ and $\tau_R$ as functions of the swimmer concentration $n$, all measured in the lab coordinate system. In all cases, results are presented for suspensions of pushers, pullers, and non-interacting swimmers, where the latter refers to simulations in which hydrodynamic interactions between swimmers have been disabled by setting all terms containing the fluid velocity $\mathbf{U}$ in their EOMs~\eqref{eq:rdot}--\eqref{eq:pdot} to zero, while keeping the tracer EOMs~\eqref{eq:rdot_tr}--\eqref{eq:pdot_tr_infAspect} unchanged; note that, in this limit, pushers and pullers are statistically equivalent~\citep{Bardfalvy:SoftMatter:2019}). Thus, at the lowest swimmer concentrations ($n < 10^{-2}$), where swimmer-swimmer correlations are small, $D_T$ and $D_R$ are both equivalent for the three swimmer types, and all show a linear increase with $n$; for $D_T$, this is in accordance with previous results from experiments~\citep{Leptos1,Mino1,Jepson1,Ortlieb1,Wu_Libchaber:PRL:2000}, simulations~\citep{Delmotte1,Stenhammar1,Krishnamurthy1}, and theory~\citep{Morozov:PRX:2020,Lin1,Pushkin1,Thiffeault1}. 

For intermediate concentrations ($0.01<n<0.2$), significant deviations from the linear dependence develop due to swimmer-swimmer correlations, with pushers showing a more steep increase, and pullers a slower increase than the noninteracting swimmers~\citep{Krishnamurthy1,Stenhammar1,Qian1}; for pusher suspensions, this corresponds to the buildup of correlations leading up to the transition to active turbulence. Above the transition ($n \geq 0.2$), the increase in both $D_T$ and $D_R$ becomes less steep, indicating a different scaling behavior inside the turbulent regime. While this behaviour for $D_T$ is similar to the behaviour of the RMS fluid velocity $\langle U^2 \rangle$ studied in~\citep{Bardfalvy:SoftMatter:2019}, the three corresponding regimes for $D_R$ are somewhat less well-defined. It is furthermore clear from Fig.~\ref{fig:D_Tau}ab that the relative effect of correlations is approximately an order of magnitude smaller for $D_R$ than for $D_T$, in accordance with the qualitative observations in Fig.~\ref{fig:overview}. 

\begin{figure*}[ht!] 
\begin{center}
\includegraphics[width=160mm]{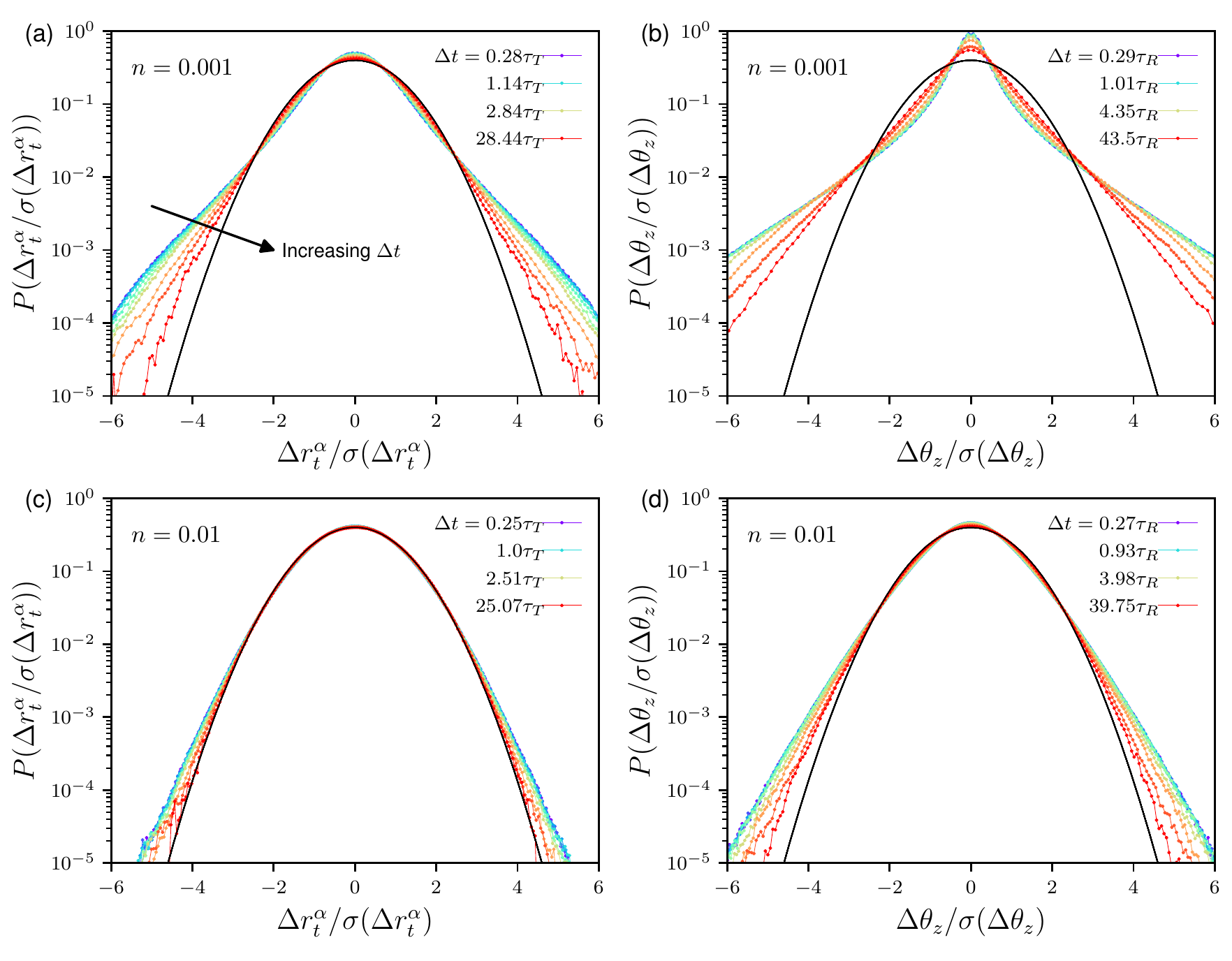}
\caption{\textbf{PDFs of tracer displacements.} Normalized PDFs of translational (a, c) and rotational (b, d) displacements for different lag times $\Delta t$ at two different densities, as indicated, where $\alpha$ denotes Cartesian components. The PDFs have been rescaled by their standard deviations $\sigma$, and times are expressed in terms of of the fitted correlation time $\tau_T$ or $\tau_R$ at the corresponding density. The solid black lines show the corresponding unit-variance Gaussian functions for reference.} 
\label{fig:PDFs}
\end{center}
\end{figure*}

Furthermore, the corresponding persistence times $\tau_T$ and $\tau_R$ (Fig.~\ref{fig:D_Tau}cd) for pushers show sharp peaks at concentrations corresponding to the transition to active turbulence ($n \approx 0.15$), with the plot for $\tau_T$ highly similar to the one for the persistence time of the fluid velocity in~\citep{Bardfalvy:SoftMatter:2019}. Notably, however, the peak in $\tau_R$ is significantly flatter and shifted towards higher densities than for $\tau_T$, highlighting the complex nature of the flow fields near the transition from disordered to collective motion. The latter peak furthermore shows values of the persistence time approximately ten times smaller than for translation, indicating that velocity gradients decorrelate faster than the velocity field itself.

To quantify the anisotropy of the translational diffusion, in Fig.~\ref{fig:anisotropy} we show the ratio $D_{\parallel}/D_{\bot}$ 
of diffusion coefficients along the tracers' major and minor axes, as described in Section~\ref{section:method}. First, we observe that suspensions of pullers and non-interacting swimmers display no measurable anisotropic diffusion, since the ratio $D_{\parallel}/D_{\perp}$ remains close to unity across the whole range of swimmer concentrations. Pusher suspensions, on the other hand, exhibit a steep increase in this ratio around $n=0.15$, before stabilizing at a value of $\sim 1.1$. This sharp increase coincides with the transition to active turbulence (\emph{c.f.} Fig.~\ref{fig:D_Tau}ab), and thus appears to be a signature of this transition. Notably, the abrupt change in the ratio $D_{\parallel}/D_{\bot}$ appears significantly sharper than the more gradually growing deviation from linear behaviour seen in Fig.~\ref{fig:D_Tau}ab. Furthermore, although the ratio $D_{\parallel}/D_{\bot}$ exceeds unity, corresponding to a higher diffusion coefficient parallel to the major axis than perpendicular to it, it is still significantly smaller than the value $D_{\parallel}/D_{\bot} = 2$ expected for Brownian diffusion~\citep{Dhont:Dynamics,Han:JCP:2010}. This is not suprising \emph{per se}, as the origin of the latter value is the anisotropic drag of an elongated particle, which does not influence the hydrodynamic advection underlying the increased active diffusion studied here. The behaviour is however qualitatively different from what was experimentally observed in~\citet{Peng1} for ellipsoidal particles immersed in a quasi-2D \emph{E. coli} suspension, where $D_{\parallel}/D_{\bot} < 1$ at densities corresponding to collective motion; we discuss this apparent discrepancy further below. 

\begin{figure}[ht!] 
\begin{center}
\includegraphics[width=80mm]{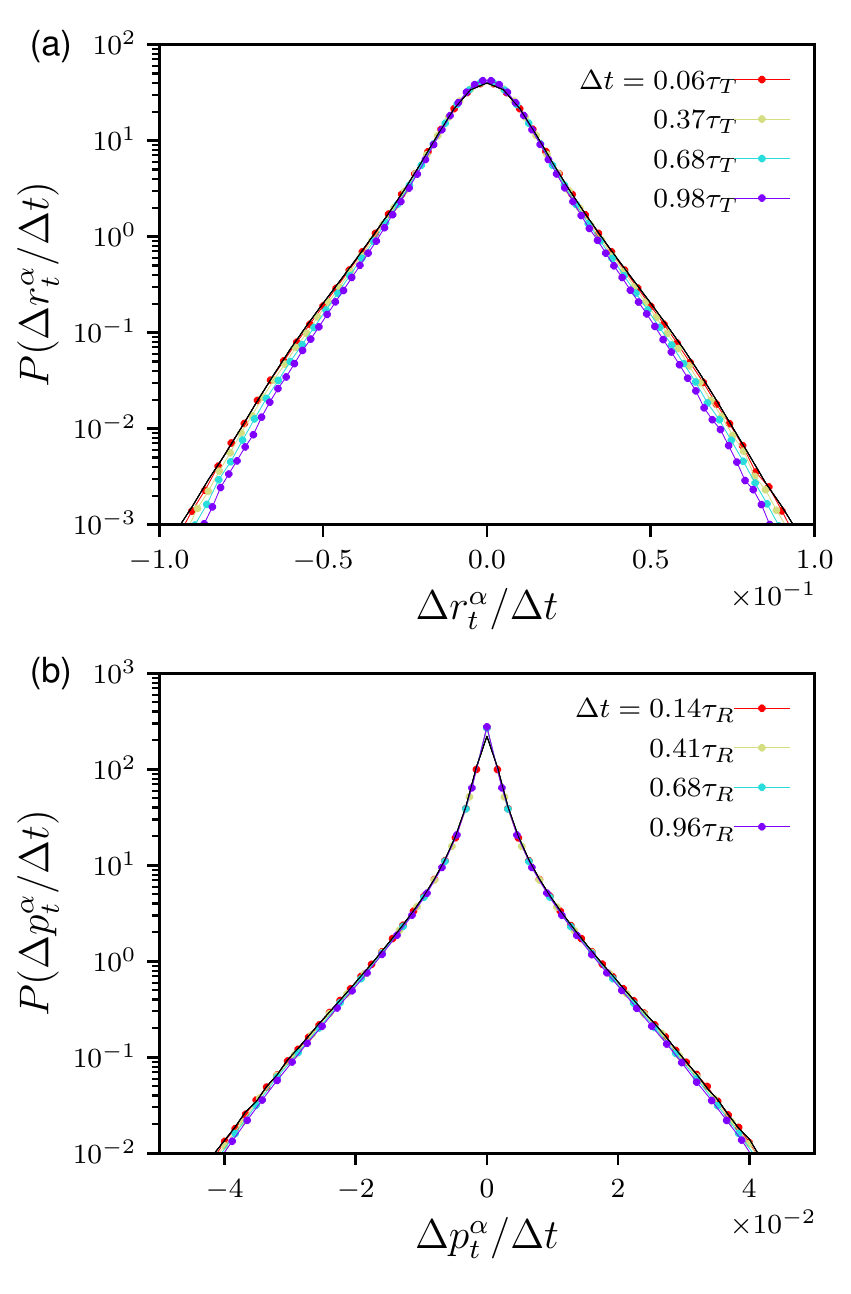}
\caption{\textbf{PDF of translational (a) and rotational (b) displacements in the short-time, ballistic regime}. In (a), the displacement data is the same as the short-time data in Fig.~\ref{fig:PDFs}a, but displacements have instead been rescaled by $\Delta t$. Note the collapse of $P(\Delta r_t^{\alpha} / \Delta t)$ to the PDF of the components of the instantaneous fluid velocity $\mathbf{U}$ (solid black line) for short $\Delta t$. Panel (b) shows the corresponding data for the rescaled orientational displacement $\Delta p^{\alpha}_t/\Delta t$ for $\Delta t < \tau_R$. The solid black line shows the PDF of the corresponding velocity gradient, as described in the text. }
\label{fig:PDFs_trescale}
\end{center}
\end{figure}

In Fig.~\ref{fig:PDFs}, we show the rescaled probability distributions (PDFs) of translational and rotational tracer displacements $\Delta r_t^\alpha$ and $\Delta\theta_z$ as a function of the observation time $\Delta t$. For short observation times and low densities, the PDFs are strongly non-Gaussian, in accordance with experimental results~\citep{Leptos1,Ortlieb1}. Theoretically, this effect can be attributed to the fact that significant displacements are driven by only a small number of swimmer-tracer encounters within the observation window, so that the central limit theorem does not apply~\citep{Kanazawa1,Zaid1,Thiffeault1}. Notably, the PDFs of rotational displacements deviate significantly more from the Gaussian form than the translational counterpart at the same density, again showing that rotational diffusion is driven by fewer swimmer-tracer scattering events than translational diffusion. This is in line with both the observed tracer trajectories (Fig.~\ref{fig:overview}a) and the relatively small enhancement of the rotational diffusion observed in Fig.~\ref{fig:D_Tau}. For short observation windows $\Delta t$, the non-Gaussian PDFs furthermore collapse when rescaled with their standard deviations $\sigma$, while, as $\Delta t$ is increased, both the translational and rotational distributions start approaching Gaussian distributions due to the growing number of swimmer-tracer scattering events within the time window. This data collapse is qualitatively in accordance with previous experimental~\citep{Leptos1,Ortlieb1} and theoretical~\citep{Thiffeault1,Lin1,Delmotte1} results, where it has been observed in both the diffusive (large $\Delta t$) and the ballistic (short $\Delta t$) regimes. In Fig.~\ref{fig:PDFs}, the collapse generally appears to occur in the regime $\Delta t < \tau_{T/R}$, \emph{i.e.}, as long as the length of the observation window is within the ballistic regime of the translational or rotational motion. This observation can be simply understood as follows: For purely ballistic translational motion, \emph{i.e.}, for $\Delta t \ll \tau_T$, the displacement $\Delta \mathbf{r}_t$ is given by $\Delta \mathbf{r}_t \approx \mathbf{U}(\mathbf{r}_t) \Delta t$. When rescaled by $\Delta t$, the PDF of the tracer displacement in the ballistic regime therefore has to collapse onto that of the fluid velocity. For the individual components of $\Delta \mathbf{r}_t$ and $\mathbf{U}$, we thus expect that $P(\Delta r^{\alpha}_t/\Delta t) = P(U_{\alpha})$, with $\alpha$ denoting Cartesian components. For rotational displacements, the same scaling argument holds for the short-time PDFs of the orientation displacement, $P(\Delta p^{\alpha}_t/\Delta t)$. As per Eq.~\eqref{eq:pdot_tr_infAspect}, these should furthermore coincide with the PDF of $(\delta_{\alpha \beta} - p_{\alpha}p_{\beta})\partial_{\gamma} U_{\beta} n_{\gamma}$, \emph{i.e.}, the components of the velocity gradient tensor projected onto randomly oriented unit vectors $\mathbf{n}$ with random positions in the simulation box. From the above reasoning, it obviously follows that the tracer displacement PDFs in the ballistic regime should collapse also when rescaled by their standard deviation, as in Fig.~\ref{fig:PDFs}. This argument shows that, for small enough $\Delta t$, the ``diffusive'' collapse of the PDFs observed in Fig.~\ref{fig:PDFs} is in fact a consequence of \emph{ballistic} particle motion at short times. This picture is verified in Fig.~\ref{fig:PDFs_trescale}: For very short times, the collapse of $P(\Delta r^{\alpha}_t/\Delta t)$ and $P(\Delta p^{\alpha}_t / \Delta t)$ for different $\Delta t$ is essentially perfect, while it gradually worsens as $\Delta t$ approaches $\tau_T$ or $\tau_R$, as expected. 

It should finally be noted that the above collapse mechanism is different from the one studied theoretically in~\citep{Thiffeault1}, which focusses on time windows within the \emph{diffusive} regime of the MSD, \emph{i.e.}, for $\Delta t > \tau_T$. While experimental studies on tracers in \emph{C. reinhardtii} suspensions~\citep{Leptos1,Ortlieb1} have focussed on the diffusive collapse, the squirmer simulations by~\citet{Lin1} showed a corresponding data collapse also in the short-time, ballistic regime. Short-time ballistic tracer MSDs have furthermore been observed in most experimental studies of tracer dynamics in \emph{E. coli} suspensions~\citep{Patteson1,Wu_Libchaber:PRL:2000,Lagarde1,Mino1}, with measured values of $\tau_T$ between 0.02 and 2 seconds. These values are comparable to our \emph{E. coli}-like system (Fig.~\ref{fig:D_Tau}c) and well within the accessible range of particle tracking experiments. Thus, verification of this ``ballistic collapse'' in \emph{E. coli} suspensions would be an interesting topic for further experimental investigations.

\section{Summary and conclusions}

In this study, we have computationally investigated the dynamics of ellipsoidal tracer particles in 3-dimensional microswimmer suspensions, using large-scale lattice Boltzmann simulations. Our model accurately incorporates far-field hydrodynamics due to dipolar flow fields which have previously been shown to dominate the enhanced diffusion of tracer particles in 3-dimensional \emph{E. coli} suspensions~\citep{Jepson1}. 

In line with previous theoretical and computational results~\citep{Stenhammar1,Morozov:PRX:2020,Qian1}, we found positive deviations from a linear dependence of $D_T$ and $D_R$ on density in pusher suspensions, coinciding with the onset of significant swimmer-swimmer correlations. Notably, $D_T$ deviates from linear behaviour by an order of magnitude more than $D_R$, indicating that the fluid velocity, which advects the particles, is significantly more affected by these correlations than the velocity gradient, which rotates them. This is in accordance with the visual nature of the tracer dynamics (see Fig.~\ref{fig:overview}a and movies in~\citep{SM}), which show particles being advected over long distances without reorienting, punctuated by short reorientation events when the tracer moves through regions with sufficiently large velocity gradients. This qualitative picture of the particle dynamics is further quantified by the fact that the rotational displacement PDFs in Figs.~\ref{fig:PDFs}bd are more strongly non-Gaussian than the translational equivalents in Fig.~\ref{fig:PDFs}ac. 

By separately evaluating diffusion coefficients parallel ($D_{\parallel}$) and perpendicular ($D_{\bot}$) to the tracer major axes, we were furthermore able to quantify the anisotropy of translational diffusion through the ratio $D_{\parallel}/D_{\bot}$. In the case of pushers, this ratio displays a sharp increase coinciding with the onset of active turbulence. In fact, this increase appears sharper than what is seen for $D_T$ or $D_R$, indicating that $D_{\parallel}/D_{\bot}$ might be used as a robust fingerprint for locating the transition to active turbulence, similar to the previous proposal by~\citet{Krishnamurthy1} who suggested to use $D_T$ as an observable sensitive to the transition to collective motion. 

As highlighted above, this behaviour is furthermore in qualitative contrast with experimental measurements on ellipsoidal tracer particles in a thin film of \emph{E. coli} suspension~\citep{Peng1}, which showed a monotonic \emph{decrease} in $D_{\parallel}/D_{\bot}$ with density. Our results show that, in unbounded 3D systems, this type of anisotropy cannot be attributed to the generic far-field hydrodynamic advection by dipolar swimmers. The hydrodynamic argument for the observed anisotropy given in~\citep{Peng1} for pushers and~\citep{Yang1} for pullers is instead based on the advection by a single swimmer moving in the same 2-dimensional plane as the tracers, while advecting and rotating them with a 3-dimensional dipolar flow field. This specific geometry breaks the 3D pusher-puller equivalence at the single-swimmer level and therefore yields different translation-rotation couplings for tracers in pusher and puller flow fields, as was shown in~\citep{Peng1}. Nevertheless, given that we observe no measurable anisotropic diffusion in 3D until near the onset of active turbulence, we argue that the far-field, single-swimmer contribution to the experimentally observed anisotropic diffusion is generally small, although it would be interesting to investigate how 2D tracer confinement affects this finding, since such confinement is known to strongly affect the hydrodynamic interactions between dipolar microswimmers~\citep{Brotto:PRL:2013,Guasto:PRL:2010,Jeanneret:PRL:2019}. Furthermore, the fact that the anomalous translation-rotation coupling in~\citep{Peng1} is only observed for high bacterial densities where collective motion occurs indicates that the single-swimmer picture is not sufficient to describe the phenomenology. Instead, we hypothesize that the anomalous anisotropic diffusion in~\citep{Peng1} is due to specific effects such as near-field hydrodynamic interactions or direct swimmer-tracer collisions that are enhanced by the quasi-2D experimental geometry, where the latter have indeed been shown to yield to $D_{\parallel}/D_{\bot} < 1$~\citep{Xu:ChinJCP:2021}. Another indication that the translation-rotation coupling is strongly affected by details of the system geometry and other experimental parameters is the contrasting results obtained for a confined, stiff colloidal chain in an \emph{E. coli} suspension, where $D_{\parallel}/D_{\bot}$ was observed to exceed unity, and even exceed the maximal Brownian value of $D_{\parallel}/D_{\bot} = 2$ for large enough microswimmer concentrations~\citep{Shafiei:SoftMatter:2020}.

In addition, we studied the probability distributions of translational and rotational displacements of tracers as functions of the length $\Delta t$ of the observation window. In accordance with previous works~\citep{Leptos1,Ortlieb1,Kanazawa1,Zaid1,Thiffeault1,Lin1,Delmotte1}, these distributions show a universal and strongly non-Gaussian behaviour, although they here only occur for short observation times. In the limit $\Delta t \rightarrow 0$, we show that this collapse is a generic consequence of the nature of ballistic particle motion: as long as particles move in straight paths, the displacement statistics must be identical when renormalized with the average length of these paths. Notably, this ``ballistic collapse'' is different from that discussed in~\citep{Thiffeault1}, where an intermittent collapse of the displacement PDFs in the diffusive regime was rationalised, showing that there are in fact two separate mechanisms leading to universal scaling of tracer PDFs in the two time regimes.

While this study forms a natural starting point for the computational investigation of anisotropic tracer motion in active suspensions, more work is needed to quantitatively connect with experimental results, which are likely to be significantly influenced by both near-field flows and non-hydrodynamic interactions between swimmers and tracers, neither of which are included in our model. An especially interesting aspect of the anisotropic tracer dynamics to be investigated in the future is the effects from changing the system geometry, which changes both the pusher-puller symmetry at low densities and affects the nature of swimmer-swimmer correlations and collective behaviour, and might therefore shift both the relative importance of rotational and translational diffusion and the nature of the translation-rotation coupling. 

\begin{acknowledgments}
Helpful discussion with Davide Marenduzzo, Cesare Nardini, Felix Roosen-Runge, and Jean-Luc Thiffeault are kindly acknowledged. This work was financed through the Knut and Alice Wallenberg Foundation (project grant KAW 2014.0052). J.S. gratefully acknowledges financial support from the Swedish Research Council (Project grant 2019-03718). All simulations were performed on resources provided by the Swedish National Infrastructure for Computing (SNIC) at LUNARC.
\end{acknowledgments}

\bibliography{refs_anisotropy.bib} 

\end{document}